\documentclass[final,5p,times]{elsarticle}

\usepackage{algorithmic}
\usepackage{subfig}

\usepackage{graphicx}
\usepackage{hyperref}

\usepackage{color}
\usepackage{ifthen}

\newboolean{showComments}
\setboolean{showComments}{true}
\newcommand{\todocmd}[1]{{\bf\textcolor{red}{#1}}}
\newcommand{\todo}[1]{\ifthenelse {\boolean{showComments}} {\todocmd{#1}} {}}
\newcommand{\todof}[1]{\ifthenelse {\boolean{showComments}} {\footnote{\todocmd{#1}}} {}}

\renewcommand{\v}[1]{\ensuremath{\mathbf{#1}}} 
 
\newcommand{\braket}[2]{\left< #1 \vphantom{#2} \right | \left. #2 \vphantom{#1} \right>} 
\let\baraccent=\= 
\renewcommand{\=}[1]{\stackrel{#1}{=}} 
\providecommand{\e}[1]{\ensuremath{\times10^{#1}}}
\journal{Computer Physics Communications}

\begin{document}

\begin{frontmatter}

\title{An OpenCL implementation for the solution of TDSE on GPU and CPU architectures}

\author{Cathal \'O Broin}
\ead{cathal.obroin4@mail.dcu.ie}

\author{L A A Nikolopoulos}
\ead{lampros.nikolopoulos@dcu.ie}

\address{School of Physical Sciences, Dublin City University, Collins Ave and~
National Center for Plasma  Science $\&$ Technology,\\ Collins Ave, Dublin 9, Ireland}

\begin{abstract}
Open Computing Language (OpenCL) is a parallel processing language that is ideally suited for running parallel algorithms on Graphical Processing Units (GPUs). In the present work we report on the development of a generic parallel single-GPU code for the numerical solution of a system of first-order ordinary differential equations (ODEs) based on the OpenCL model. We have applied the code in the case of the time-dependent Schr\"odinger equation  of atomic hydrogen in a strong laser field and studied its performance on NVIDIA and AMD GPUs against the serial performance on a CPU. We found excellent scalability and a significant speed-up of the GPU over the CPU device which tended towards a value of about 40 with significant speedups expected against multi-core CPUs.
\end{abstract}
\begin{keyword}
General Purpose Graphical Processing Unit (GPGPU) Programming \sep Taylor Series \sep Runge-Kutta methods \sep Time-Dependent Schr\"odinger Equation \sep Quantum Dynamics \sep Ordinary differential equations 
\end{keyword}
\end{frontmatter}


\section{Introduction}
\label{sec:intro}
Exploration of the fundamental processes that occur when atomic and molecular systems are subject to extreme conditions is currently a major research area.  Theoretically, it is a huge task to treat the exact time-dependent (TD) response of a multi-electron system subject to a strong electromagnetic (EM) field by \textit{ab initio} methods. In response to extensive experimental achievements using high-intensity Ti:Sapphire laser sources in the long wavelength regime, a very successfull approach that adopts the single-active-electron (SAE) approximation was applied to the atomic and molecular cases~\cite{kulander:1987b}. For systems of only two electrons, such as the negative hydrogen ion, helium and molecular hydrogen, direct ab-initio solutions of the time-dependent Schr\"{o}dinger equation (TDSE) have appeared in the early nineties (for a review see ref. \cite{lambropoulos:1998}). Since then, computational performance has increased steadily and as a result these methods have reached a high level of accuracy, efficiency and reliability, tackling successfully the very demanding theoretical problem, of single and double ionization of helium at 390 and/or 780 nm \cite{parker:2000,parker:2006}.

Recently, the realization of short-wavelength sources, through the free-electron laser of high-order harmonic generation techniques, which deliver brilliant radiation in the soft- and (in the immediate future) hard X-ray regime have initiated new challenges in the strong-fields atomic and molecular physics \cite{WabnitzEtAl2005, sorokin:2006}. Theoretical and computational approaches to tackle these challenging problems have been developed in atomic and molecular physics studies or are underway. Those include variants of time-dependent Hartree-Fock (TDHF)~\cite{kulander:1988,kulander:1991,bandrauk:2006,scrinzi:2005} as well as extensions of the traditional time-independent R-matrix method to the time-domain such as, the R-matrix Floquet approach \cite{burke:1991}, TDSE approaches fully based on R-matrix theory \cite{hart:2007,guan:2007,hart:2008} and the recently developed R-matrix incorporated time (RMT) method \cite{NikolopoulosEtAl2008, LysaghtEtAl2011, Moore2EtAl2011}. Thus it appears that there is a consistent interest in the development of computationally tractable methods able to treat multi-electron systems with the least approximations possible. 

In the past two decades there have been several advances in various directions in the computational infrastructure. When a computationally demanding problem is being tackled the entire computing environment should cooperate in a coherent manner. This includes reliable and robust numerical libraries, sophisticated compilers, high speed networks, visualization software, technical support and training together with high processing rate and fast memory.  The emergence of Central Processing Unit (CPU)-based parallel architectures allowed the development of High Performance Fortran, various parallel versions of C++ and the successful usage of \emph{Message Passing Interface} (MPI) and \emph{Open Multi-Processing} (OpenMP). As it is not the purpose of this work to elaborate on the available techniques for CPU-based computational paradigms we will focus on an alternative possibility of growing interest, namely the use of a heterogenous computational enviroment which involves the use of \emph{General Purpose Graphics Processing Units} (GPGPUs) for an efficient and low-cost \emph{distributed} hybrid computing system \cite{PennycookEtAl2011}.

The \textit{GPGPU} programming model has appeared recently due to the availability of high-level compilers, through C-like languages such as CUDA and OpenCL C as well as PGI CUDA Fortran, where commands are addressed directly to the Graphics Processing Unit (GPU) \cite{NVIDIACUDA, KhronosOpenCL, PGICUDAFortran}, FortranCL is an OpenCL fortran interface implementation which originated as an interface within the quantum chemistry project for Octopus, a TDDFT package \cite{FortranCL, MarquesEtAl2003}. This allows OpenCL functions to be called from Fortran code. The major advantage of the architecture is the large number of, what are effectively, cores present on a GPU. Thus a powerful desktop computing environment appears feasible, provided some current drawbacks are resolved such as the large disparity between double precision and single precision performance, possibly due to the lack of dedicated double precision arithmetic units, and the availability of high-performance library routines. As a result of the possibilities with the GPGPU platform, it represents a hot topic within computational physics. Two main computational platforms for GPU computing are currently in the mainstream interest; namely the CUDA and OpenCL frameworks. At the moment CUDA is of heavy use in a number of different scientific areas, but interest in OpenCL is increasing, with tools also available to convert CUDA code into OpenCL code such as the program Swan\cite{HarveyFabritiis2011}. OpenCL is a language that was designed to suit the parallelism of GPUs. It is, in essence, very similar to CUDA but in terms of features within the framework there are some significant differences arising from CUDA being limited to a particular set of hardware from a particular manafacturer whilst OpenCL is not.

The purpose of this work is two-fold. The first is to pursue the development of a theoretical method to tackle demanding computational problems in the area of complex quantum systems under intense and ultrashort radiation fields. The second is to present a computational model which is only, we believe, in its starting phase, namely the development of a parallel computational model which does not discriminate between GPU and CPU 
architectures. In this sense, the present computational model is designed and is able to run on both CPU and GPU-based systems. To this end, the computational framework that we chose is based on the OpenCL language. Though the usage of GPUs is already common within fields such as quantum chemistry \cite{StoneEtAl2010} and usage is flourishing in fields as varied as fluid dynamics and magnetohydrodynamics (see the introduction from \cite{UedaEtAl2011} and citations within) and statistical physics \cite{Weigel2011} with some usage appearing in fields such as in interacting many particle systems \cite{Kramer2011}.

Since CUDA is a more mature platform there exist routines which can optimize existing codes such as the use of an existing accelerated FFT routine such as the FFTW3 library used in Ref \cite{BaukeKeitel2011}. It is worth noting that most implementations using GPUs are on the CUDA platform. Thus for this additional reason the present OpenCL implementation represents an important contribution in this newly emerged field.

The paper is organized as follows. In Sec.~\ref{sec:opencl} a basic description of the OpenCL and GPU concepts and terminology is given. In Sec. ~\ref{sec:tdse} we formulate the physical problem under question in mathematical terms. In Sec.~\ref{sec:tdse_opencl} we present the OpenCL implementation specific to the solution of a system of ODEs, while in Sec.~\ref{sec:hydrogen_opencl} we apply our approach to the case of an hydrogen atom and present the benchmarking results. Finally we have relegated some the details about the hardware tested and the specific numerical algorithms employed in this study to the appendicies.


\section{GPU programming and the OpenCL framework}
\label{sec:opencl}

\begin{figure}[!t]
\centering
\includegraphics[width=250px]{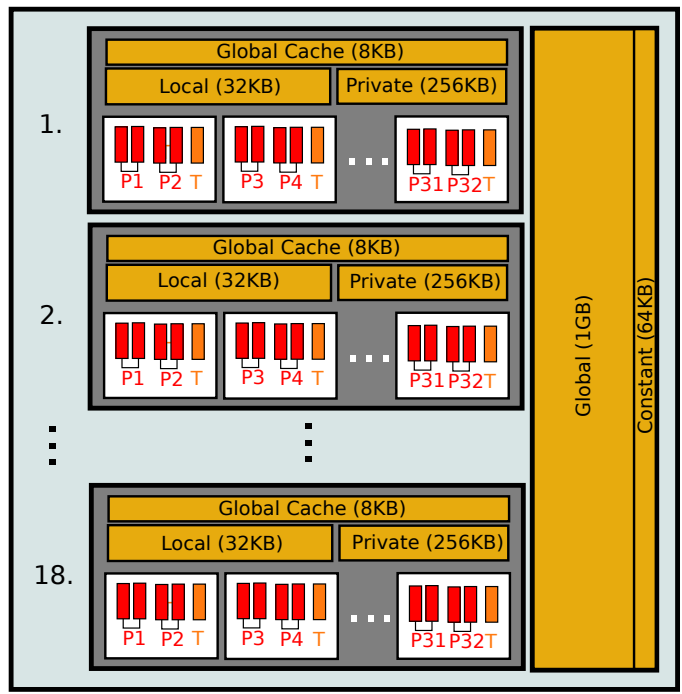}
\caption{An AMD GPU Model based on the data from \ref{ap:FirePro}. the ALU, each pair of which are grouped into a single double precision processing element, are in red, the transcendental unit, in orange and marked with the T, is not used. Each grey box represents one compute unit, of which there are 18. The pool of registers which form private memory are shared amongst processing elements in a compute unit. The global cache caches global memory for use within the compute unit, it is not accessed explicitly. Also shown within the compute unit is the local memory which is accessed explicitly and is the medium for communication within a compute unit. The global and constant memory shown are accessed by processing elements within compute units. }
\label{fig:AMDGPUModel}
\end{figure}

GPUs are a type of compute device in OpenCL terminology. GPU architectures have blocks of processing elements.
Processing elements are similar to cores except for a few key differences since they are, or effectively are, arthimetic logic units (ALUs). A processing element will have access to a certain amount of memory which it exclusively accesses. This is known as private memory. Groups of processing elements execute in a SIMD fashion and may share a common memory which can be treated as local memory as described later. These groups are known as compute units. GPU branch granularity is coarse grained because of the SIMD design. GPUs typically have a slower clock speed (700-900MHz) in comparison to CPUs.
Figure \ref{fig:AMDGPUModel} demonstrates a typical configuration for an AMD GPU. The specific architectures provided by different vendors may vary, but the abstraction provided by OpenCL will hold.

Global memory is available for access to all compute units. Constant memory is a part of global memory which is not changed by processing elements. A cache for global memory may also be available. Global Memory is typically not integrated onto the same chip as the GPU. For a CPU, RAM typically uses DDR2 and the newer DDR3 whilst a GPU typically accesses GDDR5 global memory. For AMD hardware GDDR5 memory has twice the bandwidth of DDR3 memory \cite{GDDR5}.

Communication between the GPU and CPU typically occurs over a PCI Express x16 connection. For the V7800 this gives a theoretical maximum bandwidth of 8 GB/s while the peak realizable bandwidth is 6 GB/s.

\subsection{OpenCL}
OpenCL is a royalty free open standard. Initially developed by Apple\textsuperscript{\textregistered}  Inc, the standard is being actively developed and worked upon by the Khronos group, a large multi-vendor consortium which includes companies such as NVIDIA\textsuperscript{\textregistered}, AMD\textsuperscript{\textregistered}, IBM\textsuperscript{\textregistered}, ARM\textsuperscript{\textregistered}, Intel\textsuperscript{\textregistered}, Texas Instruments\textsuperscript{\textregistered}, Sony\textsuperscript{\textregistered} and others.
The current implementations of OpenCL from Intel, AMD and NVIDIA are based on the 1.0 and 1.1 standards. The 1.2 standard was released on the 15 of November 2011.

OpenCL distinguishes between two types of code in any OpenCL program; the host code and the OpenCL C code.

All code that is written in standard programming languages such as C or Fortran can be regarded as host code. A regular program with no connection to OpenCL can be viewed as containing entirely host code for example. The host code interacts with OpenCL purely through function calls to the OpenCL library. This means that any compiler can be used to compile the host code as long as it can link against the OpenCL library.

The OpenCL C code is written in an OpenCL variant of the latest ANSI/ISO standard of C known as C99. The major differences between OpenCL C and C99 are the restrictions placed on the language. A key restriction is the lack of recursion due to GPU hardware issues and also that two or more dimensional arrays must be treated as one dimensional arrays when being used as arguments for kernel functions. Although complex numbers are supported by the C99 standard they are not implemented in OpenCL C, instead, the user can create a complex structure containing two double precision elements, it is then a relatively simple matter to define the relevant series of complex multiplication functions. This, however, is an undesirable additional step. It is preferable if optimized implementations were used implicitly such as in C99. Other restrictions are listed in the OpenCL specification \cite{Khronos2011}. The OpenCL C code is the code that will actually be performing a particular computation on a particular target such as a GPU or a CPU.

Whilst CUDA is portable amongst most operating systems, OpenCL is portable in the greater sense of not being limited to specific hardware as well as operating systems. Support is not dependent on a single vendor. Possible compute devices in OpenCL are not just limited to GPUs and CPUs, they can also include FPGAs, DSPs, the IBM\textsuperscript{\textregistered} Cell architecture and many more.

OpenCL C code will execute on any architecture but, in practice, it will require a slight code modification or possibly a partial rewrite to achieve good performance from one architecture to the next.

OpenCL C code is compiled during the runtime of the host code. The OpenCL C code can be specifically targeted to a particular instance of a problem; some aspects are known only at runtime of the host code. This information can be used at compile time for the OpenCL C code and thus the code can be optimized for that particular instance.  In this way, the compiler can take advantage of what is known at runtime of the host code. 

Memory objects such as one dimensional arrays can be created for use by the device code by function calls to the OpenCL library. A handle is returned to the object by the library which can then be used to refer to the object in future function calls.

OpenCL as a framework provides for the execution of functions known as kernels, which are written in OpenCL C. A kernel is not directly called by the host code, instead, a call to a specific kernel with specific memory objects as arguments is placed in a queue on the host device when the clEnqueueNDRangeKernel() function is called. The particular implementation of the OpenCL standard will take care of all further details. For example, the implementation will decide when to pass a particular batch of function calls queued  from the CPU to the hardware scheduler that is present on a GPU. The queue is said to be asynchronous.

Since the objective is parallelism, the aim is for multiple instances of the same kernel to be simultaneously executed with independent data so as to spread the workload. The hardware is set to assign instances of this execution, known as work items, with particular identification numbers. Three sets of identification numbers are given; the local, group and global IDs. OpenCL combines work items into work groups. The minimum size of a work group for an AMD GPU is 64 work items. This minium size is known as a wavefront in AMD terminology. For NVIDIA the minimum size is called a warp. AMD GPUs currently execute a half wavefront at a time on a compute unit for double precision instructions. The local ID of a work item represents its place within a work group. The purpose of the group ID is to represent a particular work groups position in relation to the other work groups. The global ID represents the position of a work item in relation to all other work items.

For a specific work group size $N_{Group}$ the global ID is equivalent to:
\begin{displaymath}
 ID_{Global} = ID_{Group} N_{Group} + ID_{Local}
\end{displaymath}

In this way a work item knows its place in the order of things in a manner similar to the concept of topologies in MPI. That is, there exists what can be thought of as a local topology between work items in a work group and a global topology between work groups in the domain of the problem. The local and global topologies can be one, two or three dimensional in their layout.

A work item can only communicate with other work items in the same work group. Unlike in MPI, the creation of virtual topologies is not built in by the framework though the equivalent can be implemented by an interested OpenCL C programmer. A work item communicates with other work items in it's work group through the use of local memory; low latency memory that may be dedicated to a particular compute unit or global memory which is remapped to the work group. This communication approach through memory is similar to that of OpenMP.

There is a limitation on communication between work items in different work groups; they cannot communicate with each other during the execution of a kernel. This limitation is due to the execution of the kernel by the many work items, the coherence of global memory cannot be known since the order of execution of the work items is determined by the hardware scheduler and not the programmer.

When one queues kernel calls via an in-order execution queue it can be guaranteed that at the start of a function call all work items executing a particular kernel see a consistent view of memory and so it can be said that the work items have been synchronized.

\section{TDSE projected on the stationary system's eigenstate basis}
\label{sec:tdse}
Let us consider a multielectron ($N-$electrons) atomic or molecular system described through the hamiltonian operator $\hat{H}_N(\tilde{r}_N)$ with, the system's electron's positions $\tilde{r}_{N}=({\bf r}_{1},{\bf r}_{2},..,{\bf r}_{N})$. We assume that the eigenstates of the hamiltonian, $\Phi_{\gamma}(\tilde{r}_{N})$, have been calculated as the solutions of the Schr\"{o}dinger equation:
\begin{equation}
\left( \hat{H}_N(\tilde{r}_N) -  E_\gamma\right) \Phi_{\gamma}( \tilde{r}_{N}) = 0,\\
\label{eq:schrodinger_N}
\end{equation}
where with the index $\gamma$ we represent all the quantum mechanical observables required to completely characterize the states.
Let us now consider the TDSE of the above system subject to an external time-dependent field represent by the potential operator $V(\tilde{r}_{N},t)$. The TDSE of the system in the presence of this external field is written as:

\begin{equation}
i\frac{\partial}{\partial t}\psi(\tilde{r}_{N},t)
= \left[\hat{H}(\tilde{r}_{N})+\hat{V}(\tilde{r}_{N},t)\right]\psi(\tilde{r}_{N},t),
\label{eq:tdse}
\end{equation}

supplemented with the initial condition $\psi(\tilde{r}_{N},t_0)=\psi_0$.  Thus, the main goal is to calculate the time-dependent wave function of the system, given the hamiltonian $\hat{H}_N$ of the unperturbed system and the external potential $\hat{V}(t)$.  To this end, since any physical state of a quantum mechanical system can be expanded in terms of it's eigenstates, the $N-$electron wavefunction is written as: 
\begin{equation}
\psi(\tilde{r}_{N},t)=\sum_{\gamma} \!\!\!\!\!\!\!\! \int C_\gamma(t), \Phi_{\gamma} (\tilde{r}_{N}), 
\label{eq:wavefunction_time}
\end{equation}
where $\gamma$, in principle, includes both the bound and continuum states of the spectrum. At this stage, towards developing a method of calculating the TD wavefunction, we consider the standard approach which assumes that the system is enclosed in a box. Having assumed this, the bound and the continuum 
solutions of the system can now both treated with a common indexing representing a now discretized spectrum. Formally, projection of the known discretized  channel states $\Phi_\gamma$ onto the TDSE and subsequent integration over all spatial variables $\tilde{r}_{N}$ will lead to the following set of coupled partial differential equation for the radial motion in channels $\gamma$:
\begin{eqnarray}
i\frac{d}{dt}C_{\gamma}(t) &=&
E_{\gamma}C_{\gamma}(t)+\sum_{\gamma^{\prime}}  V_{\gamma,\gamma^{\prime}}(t)C_{\gamma^{\prime}}(t), \label{eq:tdse_ode}\\
E_{\gamma} &=& \langle \Phi_{\gamma} | \hat{H}_N | \Phi_{\gamma^\prime}  \rangle  \label{eq:tdse_e}\\
V_{\gamma,\gamma^{\prime}}(t) &=& \langle \Phi_{\gamma} | \hat{V}(t) | \Phi_{\gamma^\prime} \label{eq:tdse_d}\rangle  
\end{eqnarray}
By properly ordering the coefficients $C_{\gamma}(t)$ into a column vector ${\bf C}(t)$ and the diagonal ($E_{\gamma}$) and non-diagonal ($\hat{V}_{\gamma,\gamma^{\prime}}(t)$) elements into a square matrix 
$\hat{{\bf H}}(t)$ we may rewrite the TDSE governing the system-field dynamics, in matrix representation, as,
 \begin{equation}
i{\bf \dot{C}}(t)=\hat{{\bf H}}(t){\bf C}(t),
\label{eq:tdse_ode_c}
\end{equation}
supplemented with the initial condition $C(0) = C_0$. The latter set of equations for the coefficients, which
represents a system of coupled ordinary differential equations (ODE), are subject to numerical solution. The ODE form of the TDSE, no matter which system we have, allows us to utilize our solution algorithm at a very general level. Within this eigenstate representation of the system's TD wavefunction only two kinds of dynamical quantities are required for the solution; the eigenenergies and transition matrix elements between the system's eigenstates. The key point is that all the information about the exact nature of the system, whether multi-electron or not, whether atomic or molecular, is contained in the values of the eigenenergies and the matrix elements together with the required selection rules for the transitions. It is for this reason, that any TDSE, atomic or molecular, can be formally re-expressed as a system of ODEs that the present computational algorithm is especially important, since it is designed to accept as input elements, exactly the matrix elements of $\hat{{\bf H}}$ which uniquely define coupling between the system and field.

\subsection{Atomic/molecular system in linearly polarized radiation within the dipole approximation}
\begin{figure}[!t]
\centering
\includegraphics[width=180px, height=180px]{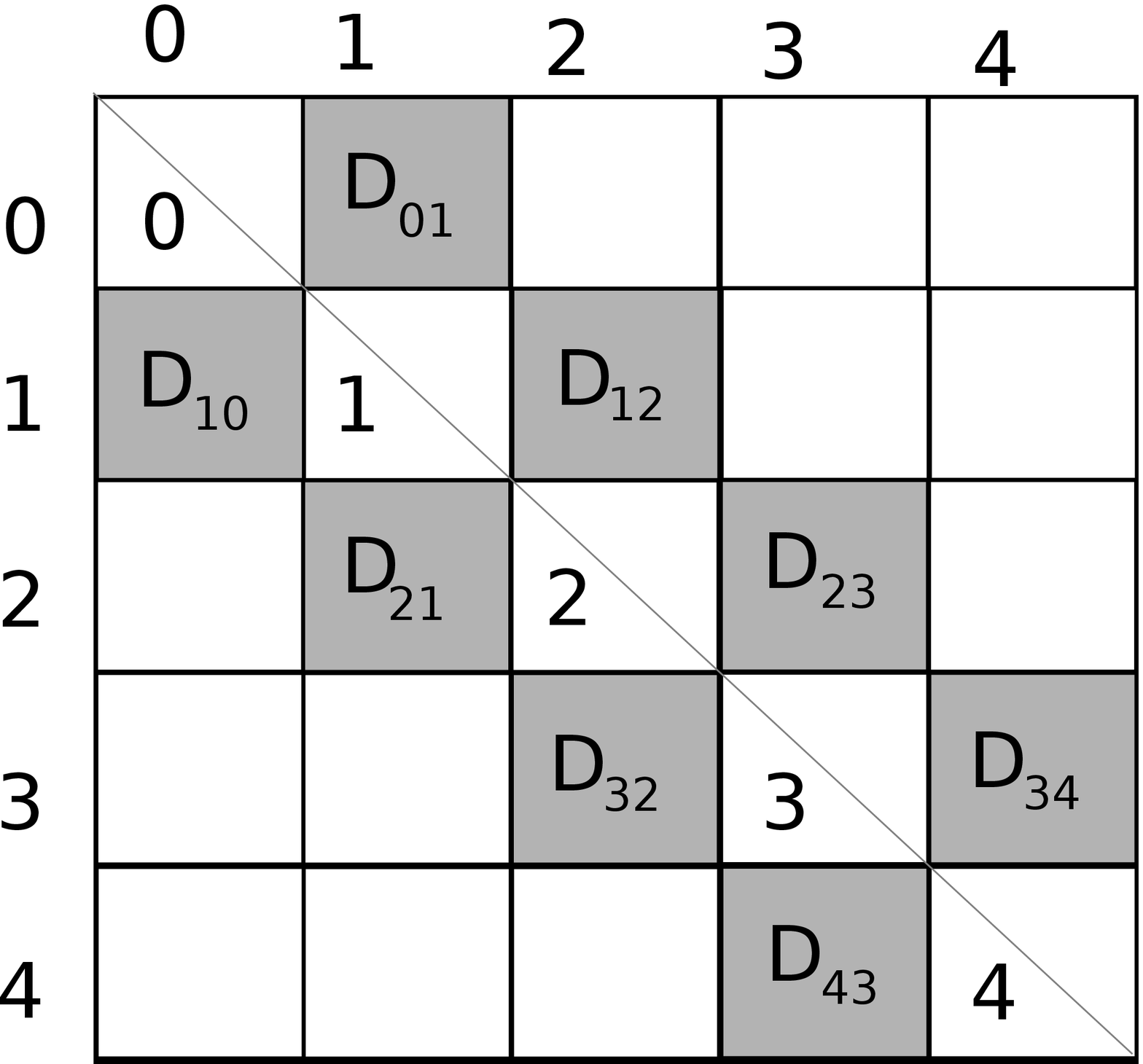}
\caption{The banded structure of the Hamiltonian is shown. This structure holds if we assume we have a linearly polarized field 
which interacts with an atomic or molecular target in the dipole approximation. The sub and super diagonal blocks contain 
$N_l$x$N_l$ transition elements. The diagonal blocks are themselves diagonal with the field-free eigenvalues for each $L$ in the diagonal.} 
\label{fig:Matrix}
\end{figure}

Since we are interested in atomic or molecular targets under EM fields one very important simplification can be achieved if we assume a linearly polarized radiation field along some axis, which without any loss of generality we may assume to be the $z-$axis in a Cartesian coordinate system. This assumption is used to determine the structure of the matrix involved in the TDSE system of equation (\ref{eq:tdse_ode_c}). Now, let us make the channel index $\gamma$ more specific; where $\gamma$ represents a solution of the hamiltonian operator. The total angular momentum quantum number, given by $\hat{L}$ with it's component along the z-axis $\hat{L_z}$ and the total spin given by $\hat{S}$ and it's component along the $z-$axis $\hat{S}_z$. Thus, we write $\gamma$ as: $\gamma = (ELSM_LM_S)$. This is the so-called $LS$ representation of the atomic states which is well suited to light atoms. 
Let us assume that the system starts from the state $\Phi_0=|E_0,L_0,S_0,M_{L_0},M_{S_0}\rangle$. It is well established, through the 
Wignet-Eckhart theorem \cite{sobelman:1972}, that in the dipole approximation for the coupling of the radiation field the non-vanishing elements are between these states with the same total spin and the same magnetic angular and spin quantum numbers. Thus we have for the dipole transition matrix elements:
\begin{eqnarray}
 D_{\gamma,\gamma^\prime} &=& \langle E_\gamma, L_\gamma, S_\gamma M_{L_\gamma} M_{S_\gamma}|
\hat{D} |E_{\gamma^\prime}, L_{\gamma^\prime}, S_{\gamma^\prime} M_{L_{\gamma^\prime}} M_{S_{\gamma^\prime}} \rangle \nonumber \\
&=& D_{\gamma,\gamma^\prime} 
\delta_{L_\gamma, L_{\gamma^\prime}\pm 1}
\delta_{S_\gamma, S_{\gamma^\prime}}
\delta_{ M_L,M_{L_{\gamma^\prime}} }
\delta_{ M_S,M_{S_{\gamma^\prime}} } 
\end{eqnarray}
Therefore we get a structure for the matrix representation of $H(t)$, which is based on very general terms, as shown in Figure \ref{fig:Matrix}.
In this figure we represent the case where the maximum total angular quantum number that was considered was four ($L_{{\gamma}_{max}} = 4$). 
The sub and super diagonal blocks contain the dipole transition elements between states having $L_{\gamma}$ and $L_{\gamma} \pm 1$. 
The diagonal blocks are themselves diagonal with the field-free eigenvalues for each $L_{\gamma}$. This banded structure 
is very general and it can be shown that this can also describe the interaction of radiation fields with molecular targets as well. For 
example in the case of diatomic molecules and in the fixed nuclei approximation the set of quantum operators required to describe the 
interaction with linearly polarized fields along the interatomic axis are the hamiltonian operator, the projections of the angular quantum number along the interatomic axis ($\Lambda$) and the spin quantum number alongside its projection along the interatomic axis. In other words the $\gamma$ channel should be represented as $\gamma = | E \Lambda, S, S_z \rangle$. 


\section{The OpenCL GPU computational framework} 
\label{sec:tdse_opencl}

Ordinary differential equations of any order can be represented as a system of coupled first-order differential equations.  The algorithms described in the present section solve this generic problem for a system of $N$ equations:
\begin{equation}
\dot{\v{C}}(t) = \v{M}(t)\v{C}(t),  
\label{eq:ode}
\end{equation}
where ${\v{C}}(t)$ is the vector containing the $N$ unknown complex coefficients and ${\bf M}(t)=-i{\bf H}(t)$, where ${\bf H}(t)$ is a $N \times N$ symmetric matrix.  At this point, it is worth noting that although the specific discussion is for the case of TDSE calculations, the framework can easily be applied to ordinary differential equations of the kind shown in Equation (\ref{eq:ode}).

Essentially the derivative is calculated through efficiently implementing the following matrix-vector multiplication:
\begin{displaymath}
-i[{\v{E}}^d +  {\v{D}}(t)]{\v{C}}
\end{displaymath}
 
Due to the neighbouring $L_{\gamma}$ states being coupled (Figure \ref{fig:Matrix}), we have a nearest-neighbour computational problem for the calculation of the matrix-vector operation. We can express the right hand side 
of Equation (\ref{eq:tdse_ode_c}) in terms of eigenstate coefficients for blocks of angular momenta $\v{C}_{L_{\gamma}}(t)$:
\begin{displaymath}
\!\!\!\!\!\!\!\!\!\!\!\!\!\!\!\!\!\hat{\v{H}}(t){\bf C}(t) =
\sum_{L_{\gamma}}  \left [ E_{L_{\gamma}}^d \v{C}_{L_{\gamma}} + D_{L_{\gamma}-1, L_{\gamma}}(t) \v{C}_{L_{\gamma}-1} + D_{L_{\gamma}+1, L_{\gamma}}(t) \v{C}_{L_{\gamma}+1} \right ], \nonumber
\end{displaymath}


where $E_{L_{\gamma}}^d$ is a diagonal matrix containing the field-free eigenvalues of the eigenstates of the angular momenta block $L_{\gamma}$ and 
$D_{L_{\gamma}-1, L_{\gamma}}$ and $D_{L_{\gamma}+1, L_{\gamma}}$ are matrices that contain the dipole matrix elements that couple the states from the $L_{\gamma}-1$ and $L_{\gamma}+1$ eigenstates to the $L_{\gamma}$ eigenstates respectively. The coupling terms are time dependent.

Since we calculate the derivative by a matrix vector calculation, the Taylor series and Runge-Kutta based methods are limited by the performance of this computation.

For the Taylor series propagator the number of synchronization points is equal to the order of the problem.
For a Runge-Kutta propagator, without specific optimizations such as those required for the Dormand-Prince algorithm, the number of synchronization points is equal to the number of stages plus one. The number of stages in a Runge-Kutta algorithm is greater than the order for methods with more than 4 stages.

Having many synchronization points per order has two major negative effects. It increases the coding complexity since the calling of the different kernel functions has to be accounted for and it also decreases the ability for optimizations to be implemented since breaks in the execution stiffle latency hiding attempts.

With the explicit Runge-Kutta methods three distinct kernels are required;
\begin{enumerate}
 \item A kernel to perform the vector sum before the derivative calculation as shown in Equation \ref{RKDeriv}
 \item A kernel to perform the derivative calculation in Equation \ref{RKDeriv}
 \item A final kernel to sum all the derivatives in Equation \ref{RKFinal}
\end{enumerate}

The Taylor series, on the other hand, requires only one kernel which performs both the derivative calculation and the addition to the solution. Since no optimizations have been implemented at present it is expected that both methods should have the same execution time if the same number of derivatives are being calculated by performing matrix vector calculations.

\subsection{Algorithm for Splitting up Generic Work} \label{sec:SplitWork}
An algorithm is necessary that splits up a workload of $N_{Work}$ units into a specific number of units given by $N_{Worker}$. A worker can be a work item or a work group.
We get the start position for a specific worker:
\begin{algorithmic}
 \STATE $Remainder \gets N_{Work} \pmod{N_{Worker}}$
 \STATE $Div \gets ^{N_{Work}}/_{N_{Worker}}$
 \STATE $Start \gets Div * ID_{Worker}$
 \IF{ $ ID_{Worker} \leq Remainder$}
  \STATE $Start \gets Start + ID_{Worker}$
 \ELSE
  \STATE $Start \gets Start + Remainder$
 \ENDIF
\end{algorithmic}
 
Now a check is performed to make sure a worker has not been assigned a value that is out of range of the available work. If this over-assignment has occured the amount of work is adjusted for the worker.
\begin{algorithmic}
 \IF{$ID_{Worker} < Remainder$}
 \STATE $i \gets 1$
 \ELSE
   \STATE $i \gets 0$
 \ENDIF
 \STATE $Div \gets MIN(N_{Work} - Start, Div + i)$
 \STATE $End = Start + WorkPerWorker$
\end{algorithmic}
This stops workers accessing unallocated memory and also stops workers from performing duplicate calculations. If an exact number of workers is chosen so that the division is assured to be correct then this is unnecessary.

\paragraph{Splitting up Work Groups Amongst Blocks of Work} \label{sec:SplitGroups}
A block of work is treated as a series of tasks that involve identical instructions being executed but with different data. Due to occupancy issues it may not be ideal to assign a full block of work to one work group. The following algorithm was created to perform this split up calculation:

\begin{algorithmic}
 \IF{$N_{Groups} < N_{Work_B}$}
 \STATE $ID_{Group_B} \gets 0$ \\ 
 \STATE $N_{Group_B} \gets 1$ \\ 
  \STATE Call main algorithm in section \ref{sec:SplitWork} \\
 \ELSE
  \STATE $N_{Group_B} \gets N_{Groups} / N_{Work_B}$
  \STATE $ID_{Group_B} \gets ID_{Group} \pmod{N_{Group_B}}$
  \STATE $Start \gets ID_{Group} /_{N_{Group_B}}$
  \STATE $End \gets MIN(Start + 1, N_{Work_B})$
 \ENDIF
\end{algorithmic}
Where $N_{Groups}$ is the number of work groups available, $ID_{Group}$ is the ID of a particular work group and $N_{Work_B}$ is the amount of blocks of work to split up. After the algorithm has finished each work group will be associated with a particular block of work which is denoted by an ID $ID_{Group_B}$. The number of work groups in the block is given by $N_{Group_B}$.

The call to the main algorithm is done where $N_{Groups}$ becomes $N_{Worker}$, $ID_{Group}$ becomes $ID_{Worker}$ and $N_{Work}$ is still used as the amount of work.

\paragraph{Splitting up a Block to Work Items} \label{sec:WorkItems}
For dividing up a block of work which we gave a Block Id of $ID_{Group_B}$ to, amongst work items in a few work groups $N_{Group_B}$, firstly we must split the work available between the work groups, so the following is done:
\begin{algorithmic}
 \STATE TotalWork $\gets N_{Group} * N_{Group_B}$
 \STATE $ ID_{Group_B} \gets ID_{Local} + N_{Group} * ID_{Block}$
 \STATE Call main algorithm in section \ref{sec:SplitWork} \\
\end{algorithmic} 

Here we assign a particular ID $ID_{Group_B}$ to each work group within the Block of work with ID $ID_{Block}$. Now we can call the main algorithm to divide the section of the block assigned to each work group to the individual work items.

\section{Application to the case of atomic hydrogen in a laser field}
\label{sec:hydrogen_opencl}
In this case the $\gamma$ channel index is replaced from the energy, the angular momentum and its component along the 
field polarization axis. We ignore the spin operators and thus we have for the eigenstates of atomic hydrogen 
$\gamma = \left (\epsilon l m_{l} \right )$, where $\epsilon$ the energy eigenvalue and $l$ the angular momentum quantum number. The continuum is discretized and together with the discrete bound states \cite{2006:30}. Then, the eigenstate basis of the field free hydrogen hamiltonian is the partial wave basis $\Phi_{\gamma}(\v{r}) = \braket{\v{r}\ }{\gamma} = \frac{1}{r} P_{\epsilon_\gamma l_\gamma}(r)Y_{l_\gamma m_{L_\gamma}}(\hat{r})$ where $Y_{lm_l}(\hat{r})$ are the spherical harmonics.
\begin{eqnarray*}
\Psi(\v{r},t) = \sum_{\gamma} C_{\gamma}(t) \Phi_{\gamma}(\v{r}).
\end{eqnarray*}
The eigenstates of atomic hydrogen are coupled to each other by a strong pulsed laser. 
This laser field couples states that differ in angular momentum by $1$ unit while the magnetic quantum number was set to zero (see Figure \ref{fig:Matrix}). We set the magnetic quantum number to zero because we assume that the initial state was the ground state of hydrogen where $l=0$ and therefore $m_l=0$ and since $m_l$ is constant we do not need to take it into account. There are $649$ eigenstates associated with each angular momentum $L$ in the basis that is used for the computations in this paper. Nine of these eigenstates are boundary states of the B-Spline basis used which are fixed at $0$; a total of $640$ states then are explicitly represented in the calculations for each angular momentum. The population of the continua represent the level of ionization after the laser field has passed.

The EM field was modelled by a sine squared pulse, linearly polarized along the 
$z-$axis:
\begin{equation}
{\bf E}(t) = \hat{z}E_0 sin^2(\frac{\omega }{2n} t) sin(\omega t)
\end{equation}
where $\omega$ is the photon frequency and $n$ is the number of cycles per pulse. The propagation was performed in the velocity gauge where the dipole 
operator is expressed as $D_v = - {\bf p}\cdot {\bf A}(t)/c$. For the velocity gauge a five point gaussian quadrature integrates the E field to give the vector potential ${\bf A}(t)=-\int_{t_0}^{t} dt^\prime {\bf E}(t^\prime)$ at each time step. The integration was found to perform as expected by comparing it to an analytical expression for a sine squared pulse with a particular photon frequency. The method works equally well for length gauge calculations where the electric field $E(t)$ gives the time dependence. 

The present GPU implementation of the Taylor and Runge-Kutta propagators (see appendix) was used for calculations in the case of atomic hydrogen.  The accuracy and precision of the propagators was verified by comparing the photoelectron spectrum (PES) of the system to a known working propagator. The propagator is based on a NAG Runge-Kutta based solver. Since above threshold ionization (ATI) has occurred the photoelectron spectrum is distinct.

In terms of the particular OpenCL implementation the splitting of a block to work groups was made where the number of blocks of work $N_{Work_B}$ mentioned in section \ref{sec:SplitGroups} corresponds to the number of angular momenta $L_{tot}$ when we are using the basis representation. $ID_{Group_B}$ is the ID for a particular angular momenta $L$. 

For division of the work initially the algorithm in section \ref{sec:SplitGroups} is called. This will assign $N_{Group_B}$ work groups to each angular momentum block of coefficients $C_L$. Following this, a call is necessary to divide the individual coefficents in $C_L$ amongst different work items. This is done through a call to the algorithm defined in section \ref{sec:WorkItems}. A choice of number of work groups and work group sizes was made such that for every coefficient there would be one corresponding work item.

In the benchmarks shown the matrix was treated as a very large one-dimensional array. Each diagonal matrix block $E_L^D$ was passed followed by the related superdiagonal dipole element block $D_L$ in row major form. In this form the superdiagonal blocks were transferred to the GPU but the subdiagonal blocks were not represented. Since the matrix is Hamiltonian the subdiagonal block is not necessary. An implementation was also made where both subdiagonal and superdiagonal blocks were present although the runtime was longer.

\subsection{Benchmarking Results}
Since OpenCL allows for both GPU and CPU execution we have benchmarked GPU execution on an AMD FirePro v7800, a single GPU on a NVIDIA Tesla S1070 Computing System node and a dual core Intel Xeon.

In comparing the Taylor propagator a specific step size and order was chosen so that for every computation it can be guaranteed that the propagator will maintain unitarity. The step size chosen does not represent the optimal choice and so should not be used in comparison to other methods. What is of interest is how the method scales as the work size is linearly increased. Since the method is nearest neighbour the computational overheard for the simulation also rises linearly. Any deviations from this linearity would be due to the limitations in the hardware or algorithms used.

By chosing to benchmark along the number of angular momenta in the basis set, the computational cost of the problem can be increased linearly by increasing the number of angular momenta. The computational cost is linear as the matrix vector calculation is, in this particular application, a nearest neighbour problem.
The size of the Hamiltonian in terms of number of double precision elements is $(N_l + N_l^2)l$ where $N_l$ is the number of pairs of double precision elements required to represent the coefficients for each angular momenta. For example, for the case of $N_l = 640$ and $l=0,...,20$ then the hamiltonian has 8615040 double precision values which require an array of about 65 MB.

An approximate comparison between the Taylor propagator and the Runge-Kutta method was made by comparing a 10th order Taylor propagator to a classic 4th order Runge-Kutta propagator and the 4th (5th) order Embedded pair Runge-Kutta-Felhberg (RKF) method. The 5th order solution was chosen from the RKF method. It has been noted in the literature that high order Taylor propagators with large step sizes perform better than lower order Taylor propagators with smaller time steps \cite{ParkerClark1996}. As a result of this a 10th order Taylor propagator was chosen. So we can compare like with like the step size of the Taylor propagator was altered so that an equal number of matrix vector calculations would be performed. Similarly, the 5th order solution from the RKF method was also performed with an adjusted time step.

\begin{figure}[!t]
\centering
\includegraphics[width=270px]{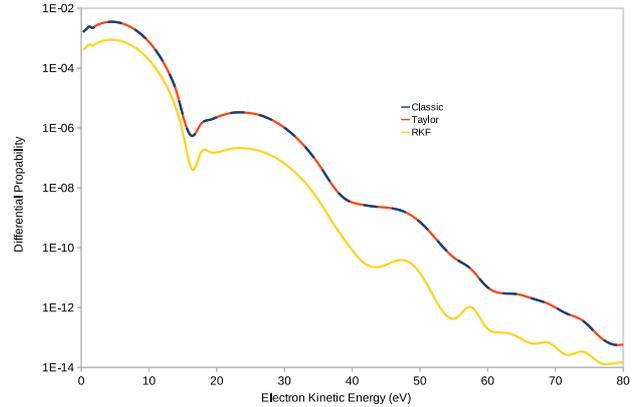}
\caption{Shown is a sample PES of simulations with 20 eV photons with a pulse of intensity $1\e{14} Wcm^{-2}$ of the three propagators that were compared. A high agreement is seen between the classic RK4 and the Taylor propagators. The Runge-Kutta-Fehlberg is markedly different from the other two methods which mostly overlap.}
\label{fig:PES}
\end{figure}

\begin{figure}[!t]
\centering
\includegraphics[width=250px]{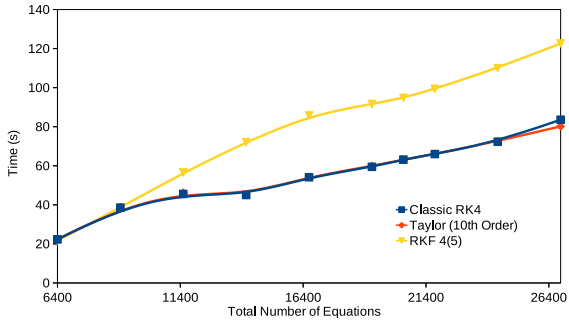}
\caption{A graph of the performance of a 10th order Taylor, 4th order Classic RK4 and 5th Order Runge-Kutta-Fehlberg propagators with time steps of $6.25\e{-3}$, $2.5\e{-3}$ and $3.75\e{-3}$ respectively. The time steps were chosen so that the same number of matrix vector calculations would be performed in all cases. Since the Fehlberg method is a multi-order method it requires slightly more computations and so does not scale as well as step-size control is not implemented. }
\label{fig:AMDMethodComparisons}
\end{figure}

As can be seen from Figure \ref{fig:AMDMethodComparisons} there was no major discrepancy in the runtime of the Taylor and Classic RK4 propagators but there was a major discrepancy with the RKF 4(5) propagator which we attribute to it being an embedded pair method. The RKF 4(5) propagator, of which we use the 5th order solution, also consistently deviates in the photoelectron spectrum, an example of which is seen in Figure \ref{fig:PES}. All three methods did contain the expected structure within the PES, but the RKF 4 (5) method deviates in the expected intensities. By comparing the unitarity of the solution of the methods a lower bound on the error can be obtained. The Taylor propagator kept unitarity to a level of $1.9\e{-14}$, while the classic RK4 method kept unitarity to $2.3\e{-11}$ and the RKF 4(5) method deviated from unitarity by $2.5\e{-11}$. Step sizes for each method were chosen so that an identical number of matrix vector calculations would be performed. The performance figures should not be used to decide on the choice of method, rather it is used here to demonstrate that the number of matrix vector calculations, which corresponds to the number of kernels queued, appears to be the primary factor for deciding the runtime speed of the algorithms.

\begin{figure}[!t]
\centering
\includegraphics[width=270px]{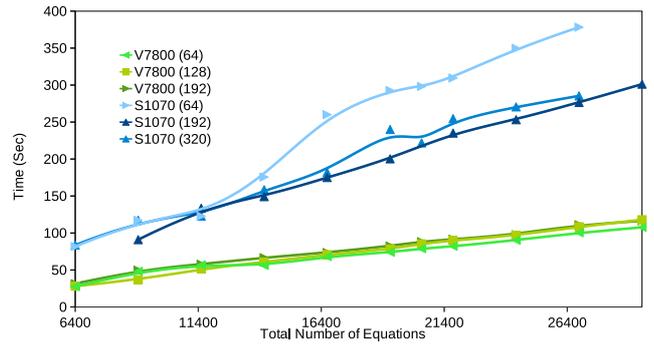}
\caption{A graph of the performance of an AMD v7800 in comparison to one GPU compute device in the Tesla S1070 for a tenth order Taylor propagator. Shown is the effects of several different configurations of work items in each work group; the work group size is shown in brackets. The optimal number of work items per work group is architecture dependent. 64 was optimal for the AMD GPU but for the NVIDIA GPU 192 work items per work group was optimal. A step size of 0.005 was chosen. The number of equations indicates the number of real equations, that is every complex equation consists of 2 real equations. } 
\label{fig:amdteslataylor}
\end{figure} 

\begin{figure}[!t]
\centering
\includegraphics[width=270px]{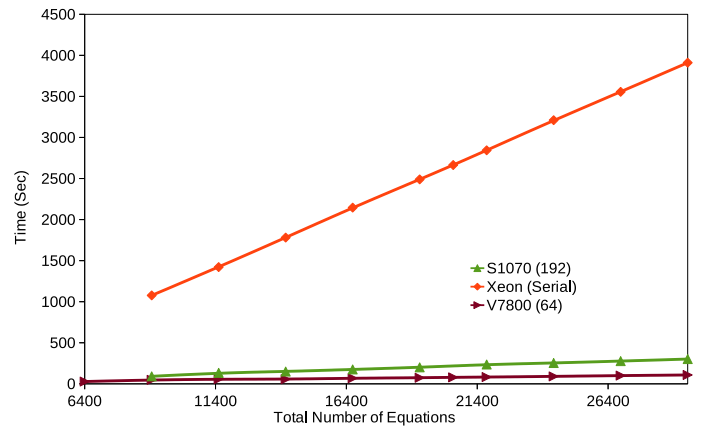}
\caption{The runtime in seconds of the best performing NVIDIA and AMD configurations from Figure \ref{fig:amdteslataylor} with an Intel Xeon.} \label{fig:CPU_AND_GPU}
\end{figure}

It can be seen from Figure \ref{fig:amdteslataylor} that for the NVIDIA GPU, 192 work items gives the greatest reduction in runtime whilst for the AMD GPU, 64 work items gives the best performance in these circumstances. The AMD GPU has a highly linear increase in run time as expected from a consistent use of the computational resources.

When the GPU results are compared to the serial CPU results a clear trend is seen (Figure \ref{fig:CPU_AND_GPU}). The GPU based simulations using OpenCL scale better than the CPUs; The runtime within the region shown in the figure for the particular pulse described, where $x$ is the number of double precision elements in the vector of coefficients, is:
\begin{eqnarray*}
& t_{INTEL}(x) = 0.14x - 170 \\
& t_{AMD}(x) = 0.0032x + 14 \\
& t_{NVIDIA}(x) = 0.010x + 9.3
\end{eqnarray*}
The CPU timing must break down for smaller situations but this is unimportant since the number of explicit states is $640$ this means the smallest possible vector of coefficients is $2560$ elements.

\begin{figure}[!t]
\centering
\includegraphics[width=270px]{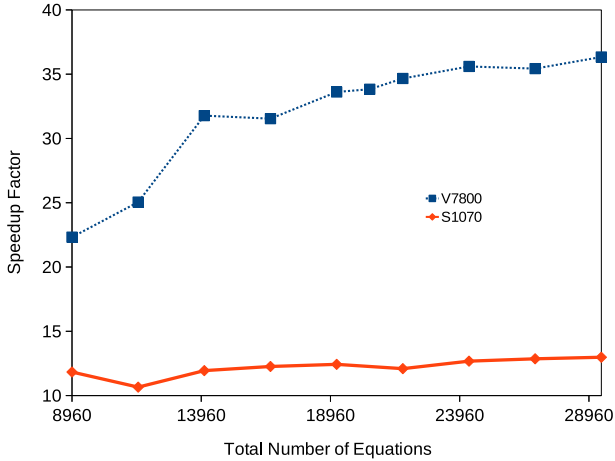}
\caption{AMD V7800 and NVIDIA T10 GPU speedups in comparison to the runtime on a single core of an Intel Xeon CPU. The speedup factor here is the ratio of runtimes $\frac{t_{CPU}}{t_{GPU}}$.} 
\label{fig:GPURUNTIME}
\end{figure}

Figure \ref{fig:GPURUNTIME} indicates the general trend which can be extrapolated from the above equations: the speedup for the AMD device tends towards a 40 times speedup whilst the NVIDIA device tends towards a 14 times speedup as the problem size increases. If, in what is most likely an overly optimistic scenario, one took the CPU scaling to be linear with the number of cores this would still provide a speedup of the GPU of an order of magnitude in comparison to a multi-core system which, incidentally, would be more expensive to purchase. With the V7800 card the relationship would terminate at a matrix which can fit into an array of size 256 MB because the largest single block of memory allocatable on the device is 256 MB.

\subsection{Optimization approaches}

Although no optimization has been performed, this represents a future line of work when the need for further runtime reductions becomes an issue. Aligning memory accesses is the first step in any optimization. It makes certain that memory accesses occur on word aligned boundaries. This is achieved by ensuring that each row of the Hamiltonian and each angular momenta block of the vector $\v{C}(t)$ are aligned.

If a large cache is available on the GPU then prefetching can be enhanced by explicit cache functions available in OpenCL. Alternatively compute unit memory can be utilized. Compute unit memory accesses have a much lower latency than global memory access by approximately an order of magnitude. Using compute unit memory also means that there will be less demand on the memory controllers.


On AMD for example, multiple accesses to a specific memory controller are serialized when there is a conflict. Atomic operations should be avoided if possible on the AMD architecture as a single atomic operation can dramatically reduce all other memory operations.
Unrolling loops can also help the compiler take advantage of the memory access structure but it also increases register use \cite{AMDProgGuide}.

Another direction to improve the present implementation is the use of 
more sophisticated propagation algorithms than the Taylor and the Runge-Kutta. An important candidate, worthy of consideration, should be the Arnoldi/Lanczos algorithm. Whilst other theoretical studies have remarked that the Taylor propagator is both simple and reliable \cite{ParkerClark1996} they argue that it is slower than the Lanczos propagator, mainly due to smaller timesteps which the Taylor propagator requires \cite{SmythCompPhysComm1998, LysaghtEtAl2011}. An optimized Taylor propagator should lend itself towards the construction of the Krylov subspace for a Lanczos propagator since the matrix vector operations represent the most significant computational bottleneck in the method.

\section{Conclusions}

A vast number of problems can be formulated in terms of a system of first-order ODEs. For propagators in which matrix vector calculations represent a significant bottleneck, significant runtime reductions can be achieved by the use of GPUs through the OpenCL language. A number of strategies for optimization exist in OpenCL which we discussed briefly for our particular case. Optimizing the existing code will require further work for an expected further order of magnitude improvement in runtime scalability. It also goes without saying that, with improvements in compilers and hardware, future trends should be for fine-tuning optimizations to be performed by sophisticated compilers and hidden behind generic functions.

\section{Acknowledgements}
This work has been funded under a Seventh Framework Programme under the project HPCAMO/256601. It has been performed in association with SCI-SYM center of DCU. The authors acknowledge 'Ireland's High Performance Computing Center  (ICHEC) for their support in using ICHEC's computational resources. Both authors acknowledge support from the COST CM0702 action. LAAN is supported by the Science Foundation Ireland (SFI) Stokes Lectureship program. 

\appendix
\section{OpenCL compute devices tested}

\paragraph{AMD FirePro V7800} \label{ap:FirePro}
The AMD FirePro V7800 is a PCI-e x16 connected graphics card \cite{AMDFIREPROV7800} with 288 processing cores. Each core consists of four arithmetic logic units (ALU) and a transcendental unit which are fed instructions through a Very Long Instruction Word (VLIW). The ALUs can be thought in OpenCL terms as a processing element. For double precision the transcendental unit is not used and the remaining four are grouped into two double precision execution units, thus there are two double precision processing elements per processing core. This means that for practical purposes 576 double precision instructions can be executed simultaneously. For floating point calculations 1152 instructions can be executed. The processing cores are grouped into compute units. Obviously the actual number of instructions executed in a cycle is dependent on the form of the workload. A compute unit (a SIMD processor) consists of 16 of the processing cores; as a result there are 18 compute units. 1 Gigabyte of global memory is available as well as 32KB of memory per compute unit. Each processing element has access to a pool of registers (256KB per compute unit). Global memory is accessed with GDDR5. The core clock is 700 MHz.

\paragraph{NVIDIA Tesla S1070 Computing Systems}
The NVIDIA Tesla S1070 computing system consists of multiple Tesla T10 GPUs which are based on the GeForceGTX 200 GPU \cite{S1070TeslaCompute2010}.
Each GPU contains 240 scalar processing cores and 4GB of memory \cite{S1070Datasheet}. 
Currently a single GPU is targeted.

\paragraph{Intel Xeon}
The Intel\textsuperscript{\textregistered} Xeon\textsuperscript{\textregistered} W3503 \cite{IntelXeonW3503} used is a 64 bit dual core CPU with a clock speed of 2.4 GHz, a 4 MB cache and with support for DDR3 memory with a 25.6 GB/s memory bandwidth.

\section{The Taylor and Runge-Kutta time propagators}

\paragraph{The Truncated Taylor Series Propagator}

Within this single-step algorithm the forwarded solution is obtained as:
\begin{equation}
{\bf C}(t+dt) = \sum_{n=0} \frac{dt^n}{n!}{\bf C}^{(n)}(t),
\label{eq:taylor}
\end{equation}
where $C^{(n)}(t)$ is the $n-$th derivative of $C(t)$ at time $t$. A recursive expression for the required $n$-th derivatives of the coefficient vector can be retrieved by successive integrations of Equation (\ref{eq:tdse_ode}) as:
\begin{equation}
{\bf C}^{(n)}(t) =\frac{-i}{n}{\bf H} (t) {\bf C}^{(n)}(t) 
\end{equation}
where the zero derivative is equal to ${\bf C}(t)$. To arrive at this expression one shall assume the time derivative of the Hamiltonian itself, within the forwarded time interval $dt$, is much smaller than the rate of change of the coefficients. Particularly for the present problem, this latter assumption is an excellent approximation, provided that the chosen time step $dt$ is much smaller the field's period  $2\pi/\omega$, in other words, $dt << 2\pi/\omega$.  It can be shown that this expression does indeed give an approximate form of the unitary operator.
In practical calculations, the above expression consists of a Taylor series truncated to some order $N$, which in combination with the time step $dt$ sets the order of accuracy of the solution. Finally, from this expression, after calculation of the derivatives of the coefficient at a known time $t$, they are combined together in a summation in order to then calculate the wavefunction at a later time $t + dt$ according to Equation \ref{eq:taylor}. The calculation for each step consists of $N$ matrix-vector multiplications and $N$ vector additions onto the solution of the system a step later.

\paragraph{Explicit Runge-Kutta Methods}
Runge-Kutta methods have been some of the most widely used methods for solving ordinary differential Equations \cite{CashKarp1990}.  Runge-Kutta methods use information from several steps to approximate a Taylor expansion \cite[pg 906]{NumericalRecipes2007}. The class of explicit Runge-Kutta methods is expressed in the form  \cite{AscherPetzol1998}:

\begin{eqnarray}
\textbf{C}(t+dt) &=& \textbf{C}(t) + dt \displaystyle \sum \limits_{i=1}^S b_{i} \textbf{C}^{(i)} \label{RKFinal} \\
\textbf{C}^{(i)} &=& \textbf{f} \left ( t + c_i dt, \textbf{C}(t) + dt \displaystyle \sum \limits_{j=1}^{i-1} a_{ij} \textbf{C}^{(j)} \right )  \label{RKDeriv}
\end{eqnarray}
where $\textbf{f} \left ( \right )$ represents the derivative of ${\bf C}$.


A number of Runge-Kutta methods exist such as the classic fourth order Runge-Kutta method and the 4th(5th) 
order embedded pair Fehlberg method\cite{Fehlberg1969}. In the Runge-Kutta-Fehlberg method 4th and 5th order steps are calculated using 
the same derivative calculation information; the difference between the two methods gives an indication of the local error size.


\bibliographystyle{elsarticle-num}
\bibliography{BibTeXMasterRecord}

\end{document}